\documentclass[%
 preprint,
 amsmath,amssymb,
 aps,
prl,
]{revtex4-2}

\usepackage{graphicx}
\usepackage{dcolumn}
\usepackage{bm}

\usepackage{siunitx}

\begin{document}


\section{Title}
Valley Polarization-Electric Dipole Interference and Nonlinear Chiral Selection Rules in Monolayer {WSe}$_2$

\section{Author list}
Paul Herrmann$^1$, Sebastian Klimmer$^{1,2}$, Till Weickhardt$^1$, Anastasios Papavasileiou$^3$, Kseniia Mosina$^3$, Zden\v{e}k Sofer$^3$, Ioannis Paradisanos$^4$, Daniil Kartashov$^{5,6}$ and Giancarlo Soavi$^{1,6,\star}$

\section{Affiliations}
\noindent
$^1$Institute of Solid State Physics, Friedrich Schiller University Jena, Helmholtzweg 5, 07743 Jena, Germany
\newline
$^2$ARC Centre of Excellence for Transformative Meta-Optical Systems, Department of Electronic Materials Engineering, Research School of Physics, The Australian National University, Canberra, ACT, 2601, Australia
\newline
$^3$Department of Inorganic Chemistry, University of Chemistry and Technology, Technicka 5, Prague, 166 28 Czech Republic
\newline
$^4$Institute of Electronic Structure and Laser, Foundation for Research and Technology, N. Plastira 100, Vassilika Vouton, 70013 Heraklion, Crete, Greece
\newline
$^5$Institute of Optics and Quantum Electronics, Friedrich Schiller University Jena, Max-Wien-Platz 1, 07743 Jena, Germany
\newline
$^6$Abbe Center of Photonics, Friedrich Schiller University Jena, Albert-Einstein-Straße 6, 07745 Jena, Germany
\newline
$^{\star}$ giancarlo.soavi@uni-jena.de

\maketitle

\section{Abstract}
In monolayer transition metal dichalcogenides time-reversal symmetry, combined with space-inversion symmetry, defines the spin-valley degree of freedom. As such, engineering and control of time-reversal symmetry by optical or magnetic fields constitutes the foundation of valleytronics. Here, we propose a new approach for the detection of broken time-reversal symmetry and valley polarization in monolayer {WSe}$_2$ based on second harmonic generation. Our method can selectively and simultaneously generate and detect a valley polarization at the $\pm K$ valleys of transition metal dichalcogenides at room temperature. Furthermore, it allows to measure the interference between the real and imaginary parts of the intrinsic (electric dipole) and valley terms of the second order nonlinear susceptibility. This work demonstrates the potential and unique capabilities of nonlinear optics as a probe of broken time-reversal symmetry and as a tool for ultrafast and non-destructive valleytronic operations.  

\section{Main text}
Time-reversal (TR) symmetry underlies some of the most exotic phases of condensed matter, including topological insulators and superconductors \cite{Sato_2017}. In monolayer transition metal dichalcogenides (TMDs), the interplay between space inversion and TR symmetry further defines the valley degree of freedom \cite{C4CS00301B,https://doi.org/10.1002/smll.201801483}, where direct transitions in momentum space at the $\pm K$ points of the Brillouin zone are energetically degenerate but non-equivalent. Engineering of TR symmetry in TMDs naturally leads to the field of valleytronics, where the degeneracy of the $\pm K$ valleys is lifted either by magnetic fields (Zeeman splitting) \cite{PhysRevLett.114.037401} or with circularly polarized light. The latter approach can be further distinguished between the generation of a real exciton population in one of the valleys \textit{via} one- \cite{Mak2012} or two-photon \cite{PhysRevLett.114.097403} absorption, or by transient breaking of TR symmetry with coherent processes such as the optical Stark and Bloch-Siegert effects \cite{doi:10.1126/science.1258122,doi:10.1126/science.aal2241}. However, in the vast majority of studies the detection of broken TR symmetry and the consequent valley polarization (VP) has been limited to the realm of linear optics, mainly the detection of a polarized photoluminescence (PL) to probe the VP induced by a real excited state population \cite{Mak2012,Zeng2012} or the detection of the Kerr rotation in a pump-probe configuration to probe valley polarized resident carriers \cite{Yang2015,Hsu2015} or valley selective coherent states \cite{doi:10.1126/science.1258122,doi:10.1126/science.aal2241}. Both approaches suffer from severe limitations: PL is intrinsically destructive, as it requires recombination of the electron-hole pair and thus the loss of the valley information while optical Kerr rotation uses a relatively intense and resonant probe pulse (\textit{e.g.}, \SI{100}{\micro\watt} of average power in Ref. \cite{Hsu2015}), which can significantly perturb the sample under investigation. In addition, both helicity-resolved PL and optical Kerr rotation can only probe the amplitude of the valley imbalance, while they do not measure the complex nature (real and imaginary parts) of the VP induced elements in the TMD susceptibility tensor. Finally, it is worth noting that both methods require low temperatures to increase the spin relaxation times \cite{10.1063/5.0002396, https://doi.org/10.1002/pssb.201552211} and thus induce a measurable degree of VP. 

In this context, nonlinear optics (NLO) can provide distinct advantages. An all-optical probe of broken TR symmetry based on NLO has been realized in layered \cite{Sun2019} and bulk magnets \cite{fiebig1994second}, and very recently in various non-magnetic TMDs under the effect of an external magnetic field  \cite{Wu2023Extrinsic}. Also in the context of valleytronics, few theoretical \cite{PhysRevB.91.041404,Cheng:19,Hipolito_2017} and experimental \cite{Herrmann2023Nonlinear,Ho2020Measuring,Mouchliadis2021} studies have recently demonstrated the advantages of a detection scheme based on second harmonic generation (SHG). All these studies were based on the measurement of a rotation in the SHG polarization ellipse while simultaneously writing the valley state with an elliptically polarized fundamental beam (FB) \cite{Herrmann2023Nonlinear,Ho2020Measuring,Mouchliadis2021}. On one hand, this approach clearly surpasses the standard methods based on polarized PL and optical Kerr rotation, because SHG is a parametric process and thus ultrafast and non-destructive, especially under the condition where the SH signal at 2$\omega$ is resonant with the exciton transition under investigation, and thus the TMD is fully transparent to the FB at $\omega$. On the other hand, detection of the VP based on elliptical SHG fails if the polarization of the FB approaches the circular state (which is the most efficient condition for the generation and detection of the VP), because in this case there is no well-defined ellipse rotation to measure. In addition, measurements of the valley SHG with elliptically polarized light are based on the assumption that the VP and electric dipole (ED, otherwise called "intrinsic") terms of the $\chi^{(2)}$ tensor are in-phase, and thus the SH rotation angle is directly proportional to the ratio $\frac{|\chi^{(2)}_{VP}|}{|\chi^{(2)}_{ED}|}$ \cite{Herrmann2023Nonlinear,Ho2020Measuring}. This, again, limits the study of broken TR symmetry in TMDs to the amplitude of the VP tensor, rather than its complex nature. As we will show, this assumption fails in the energy region of excitonic resonances, which are the ideal probe for the VP. 

In this work, we propose a new approach for all-optical detection of broken TR symmetry and nonlinear valleytronics where we simultaneously generate the VP by an off-resonant, circularly polarized FB using the optical Stark effect, and read it by measuring the resonant SH intensity rather than the polarization rotation angle. This greatly simplifies the detection scheme and enables ultrafast write/read of the VP at ambient temperature. In particular, we measure the ratio between the SH signal emitted for incoming circular \textit{versus} linear FB polarization and show that this directly probes the nonlinear elements of the $\chi^{(2)}$ tensor induced by the VP. We further demonstrate that such measurement can also probe the VP dispersion and the wavelength dependent relative phase between the VP and ED elements of the $\chi^{(2)}$ tensor. Based on this, we measure both constructive and destructive SH interference between the VP and ED terms, similar to the SH magnetic-electric dipole interference observed in bulk magnets \cite{fiebig1994second,toyoda2021nonreciprocal}. This provides a further piece of evidence for the analogies between the VP in TMDs and the magnetic-dipole response of magnets \cite{doi:10.1126/science.1258122}, as both are ultimately connected to the more general property of broken TR symmetry. Besides and beyond the scientific interest, a deeper understanding of the VP and ED nonlinear response of TMDs is of paramount importance for the development of the emerging field of nonlinear valleytronics \cite{Herrmann2023Nonlinear}.   

\begin{figure*}
\centering
\includegraphics[]{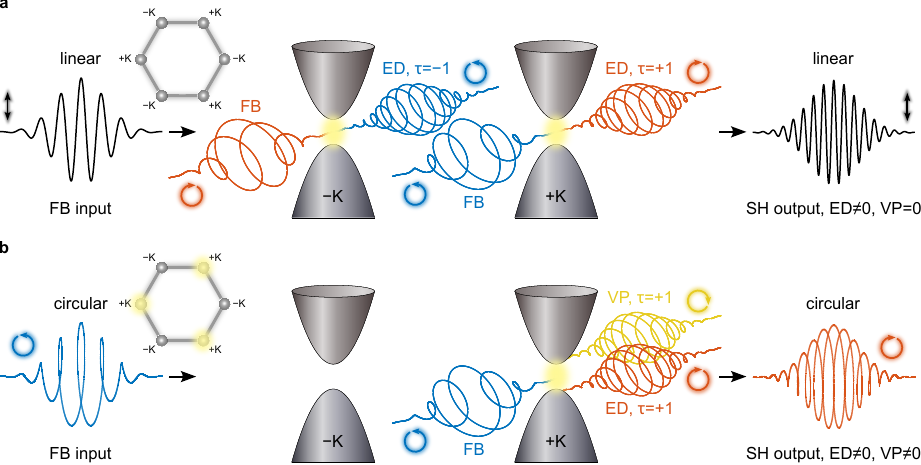}%
\caption{\label{fig:1}\textbf{Nonlinear selection rules and TR symmetry-breaking in monolayer TMDs} \textbf{a}, A linearly polarized FB (left, black) can be decomposed into right (red) and left (blue) circular components which interact with the $\mp K$ valleys respectively, emitting counter-rotating SH beams. As no VP is induced, only the ED contributes to the SH. Coherent superposition of the SH contributions from the  $\mp K$ valleys results in linearly polarized SH (black, right). \textbf{b}, A left circularly polarized FB (left, blue) interacts only with the $+K$ valley. Simultaneously, the fundamental induces a VP second order response. Therefore, in addition to the ED (red), also the VP (yellow) contributes to the counter-rotating SH. Coherent superposition of the ED and VP contributions from the  $+K$ valley results in circularly polarized SH (orange, right).}
\end{figure*}

\subsection{Crystal symmetry and nonlinear chiral selection rules}

The vast majority of NLO experiments on TMDs \cite{Dogadov2022Parametric}, such as the measurements of crystal orientation \cite{PhysRevB.87.201401}, number of layers \cite{doi:10.1021/nl401561r}, strain \cite{Mennel2018}, ultrafast switching \cite{Klimmer2021} \textit{etc.}, are based on the assumption that monolayers belong to the point group (or more precisely the \textit{wave vector group} \cite{dresselhaus2007group}) $D_{3h}$. However, a closer look shows that the wave vector group is $D_{3h}$ only at the $\Gamma$ point of the Brillouin zone, while it is $C_{3h}$ at the $\pm K$ points \cite{fajardo2019effective}. Thus, resonant excitation of the valleys should be more precisely described by the nonlinear elements of the cyclic $C_{3h}$ tensor, rather than those of the dihedral $D_{3h}$ group. The elements of the second order susceptibility $\chi^{(2)}$ for the $C_{3h}$ point group can be divided into two sub-groups, namely $\chi^{(2)}_{xxx}=-\chi^{(2)}_{xyy}=-\chi^{(2)}_{yyx}=-\chi^{(2)}_{yxy}$ and $\chi^{(2)}_{yyy}=-\chi^{(2)}_{yxx}=-\chi^{(2)}_{xxy}=-\chi^{(2)}_{xyx}$, where $x(y)$ refers to the armchair(zig-zag) axis of the crystal in the case of TMDs. The first subset is identical to the $D_{3h}$ point group and we will refer to it as the ED (or intrinsic) response ($\chi^{(2)}_{ED}=\chi^{(2)}_{xxx}$). These elements can fully describe SHG in TMDs in the case of non-resonant excitation (\textit{e.g.}, below-gap virtual states), and thus they represent the crystal (\textit{i.e.}, geometrical, intrinsic) response of TMDs. In contrast, the second subset appears only in the $C_{3h}$ group and must be taken into account in the case of resonant excitation at $\pm K$. In this regard, the second subset is a direct probe of broken TR symmetry and thus of the VP ($\chi^{(2)}_{VP}=\chi^{(2)}_{yyy}$). However, in contrast to a standard $C_{3h}$ system, the VP elements of the $\chi^{(2)}$ tensor in TMDs are also chiral, as they must describe the broken space inversion while preserving TR symmetry \cite{PhysRevLett.108.196802}. This has been observed, for instance, as a rotation angle in opposite directions in recent experiments based on SH by an elliptically polarized FB \cite{Herrmann2023Nonlinear,Ho2020Measuring}. Thus, we can further define $\chi^{(2)}_{VP}=\tau\cdot\chi^{(2)}_{yyy}$, where $\tau=\pm1$ at $\pm K$. This is the nonlinear analogous of chiral selection rules for absorption and emission of light in TMDs \cite{PhysRevLett.108.196802,C4CS00301B}. These observations have important consequences, as we show schematically in Fig.\,\ref{fig:1}. If we focus our attention only on resonant excitation at $\pm K$, we can immediately understand that linear and circular polarization of the FB will probe different symmetries. In particular, in the case of linear excitation we coherently add up the left and right circular components of the FB, while preserving TR symmetry and not producing any VP, leading to a SH signal described only by the $\chi^{(2)}_{ED}$ terms (Fig.\,\ref{fig:1}a). Thus, the effective $\chi^{(2)}$ tensor in the case of linear excitation is identical to that of the $D_{3h}$ point group and therefore it probes only the ED response of TMDs (as long as no VP is introduced by any other means). In contrast, circular polarization of the FB will simultaneously induce and probe the chiral VP elements of the $C_{3h}$ tensor, and their interference with the ED-SH signal (Fig.\,\ref{fig:1}b). As we demonstrate in the next section, this fundamental difference can be measured experimentally as the ratio of the SH intensity in the cases of circular \textit{versus} linear polarization of the FB.

\subsection{Second harmonic intensity with linear and circular polarization}

Based on the previous discussion, we can write the expression of the second order polarization $\boldsymbol{P^{(2)}(2\omega)}$ in the two cases of linear and circular FB. For linear excitation, the SH response reads (see Supplementary Information S3.1): 

\begin{equation} \label{eq:1}
    \boldsymbol{P^{(2)}(2\omega)}
    =
    \begin{pmatrix}
    P_x^{(2)}\\
    P_y^{(2)}
    \end{pmatrix}
    = \epsilon_0
    \begin{pmatrix}
    \chi_{ED}^{(2)}(E_x^2-E_y^2)\\
    - 2 \chi_{ED}^{(2)} E_x E_y
    \end{pmatrix}
\end{equation}

as expected from a system with $D_{3h}$ symmetry. Instead, for circular excitation the polarization is (see Supplementary Information S3.2):

\begin{equation} \label{eq:2}
    \boldsymbol{P^{(2)}(2\omega)}
    = 
    \begin{pmatrix}
    P_+^{(2)}\\
    P_-^{(2)}
    \end{pmatrix}
    = \epsilon_0 \sqrt{2}
    \begin{pmatrix}
    (\chi^{(2)}_{ED}+i \chi^{(2)}_{VP})E^2_{-}\\
    (\chi^{(2)}_{ED}+i \chi^{(2)}_{VP})E^2_{+}
    \end{pmatrix}
\end{equation}

where $\boldsymbol{P^{(2)}_\pm} = P^{(2)}_\pm \boldsymbol{\sigma_{\pm}}$ and $\boldsymbol{E_\pm} = E_\pm \boldsymbol{\sigma_{\pm}}$ define left and right circular polarization of the second order polarization and of the fundamental electric field, with $\boldsymbol{\sigma_{\pm}}  = \frac{1}{\sqrt{2}} ( \boldsymbol{e_x} \pm i \boldsymbol{e_y} )$. Equation\,(\ref{eq:2}) shows that for circular excitation, the SH polarization is always cross-polarized with respect to the FB \cite{Säynätjoki2017}, as imposed from the conservation of the angular momentum in NLO processes \cite{Bloembergen:80}. It is very important to highlight that this property has nothing to do with the valley degree of freedom, in contrast to the discussion of seminal reports on valley selection rules for SHG in TMDs \cite{Seyler2015}. One can easily appreciate this by setting $\chi^{(2)}_{VP} = 0$ in equation\,(\ref{eq:2}), and obtain the same result of cross-polarization between FB and SH. In addition, we note that equation\,(\ref{eq:2}) reminds of equation\,(4) from Ref. \cite{fiebig1994second} with two main differences: (1) the magnetic-dipole contribution is here substituted by the VP term; (2) the pre-factors to $E^2_{\pm}$ are identical in our case, while they have opposite sign for the magnetic-dipole term in Ref. \cite{fiebig1994second}. The latter observation derives from the different sign of $\tau = \pm 1$  for light of opposite helicity (see Supplementary Information S3.2), namely what we previously defined as the nonlinear chiral selection rule. This also has important consequences on the VP-ED interference, as we discuss in detail in the following.     

Based on equations\,(\ref{eq:1}) and (\ref{eq:2}), the VP can be measured by looking at the ratio $\eta$ of the SH intensity in the two cases of circular and linear FB polarization (see Supplementary Information S3.3):

\begin{equation} \label{eq:3}
    \eta := \frac{I_{circ} (2\omega)}{I_{lin} (2\omega)} = 2\left[1 + \frac{|\chi^{(2)}_{VP}|^2}{|\chi^{(2)}_{ED}|^2} \right]
\end{equation}

assuming equal intensity of the incident linear and circular polarized waves: $|I_0^{lin}(\omega)| = |I_0^{circ}(\omega)|$. Note that equation\,(\ref{eq:3}) is valid only in the simplest case where we neglect the complex nature of the nonlinear tensor $\chi^{(2)}$. However, already under this simplified assumption, we can immediately observe that the aforementioned ratio is exactly 2 only in the absence of a VP ($\chi^{(2)}_{VP} = 0$), namely when the system maintains the $D_{3h}$ symmetry also for circular excitation, as reported for instance in Ref. \cite{Säynätjoki2017}. However, in presence of VP, equation\,(\ref{eq:3}) predicts a ratio larger than 2, since linear excitation probes the $D_{3h}$ symmetry while circular excitation probes the broken TR symmetry $C_{3h}$.

The result becomes even more interesting if we now consider the complex nature of the nonlinear tensor $\chi^{(2)}$ \cite{fiebig1994second}. In this case, equation\,(\ref{eq:3}) becomes (see Supplementary Information S3.3):

\begin{equation} \label{eq:4}
    \eta = 2 \cdot \left[1 + \frac{|\chi^{(2)}_{VP}|^2}{|\chi^{(2)}_{ED}|^2} + \frac{|\chi^{(2)}_{VP}|}{|\chi^{(2)}_{ED}|} \sin{\Delta\varphi} \right].
\end{equation}

where we defined the complex elements of the second order nonlinear tensor as  $\chi_{ED/VP}^{(2)}=|\chi_{ED/VP}^{(2)}| \cdot e^{i \varphi_{ED/VP}}$, and thus the last term $\sin{\Delta\varphi}$ (with $\Delta\varphi = \varphi_{ED}-\varphi_{VP}$) describes the interference between the VP and ED terms of the SH signal and it can lead to a ratio larger or smaller than 2. Note that the interference is maximum if, in a specific wavelength range, one of the terms is real while the other is imaginary (namely a phase shift of $\Delta\varphi = \pm \pi/2$), as discussed in the case of SH magnetic-electric dipole interference in bulk magnets \cite{fiebig1994second}. Constructive and destructive interference at $\pm \pi/2$ (rather than 0 and $\pi$) occurs due to the intrinsic phase shift of $\pi/2$ between the SH originating from $|\chi^{(2)}_{ED}|$ and $|\chi^{(2)}_{VP}|$, as it can be observed already in equation\,(\ref{eq:2}). However, there is also one major difference compared to the results reported in the case of bulk magnets \cite{fiebig1994second}, namely the fact that in TMDs left and right circular polarization lead to the same SH intensity (note the symmetric ratio for LCP and RCP in Fig.\,\ref{fig:2}a), as a consequence of the nonlinear chiral selection rules. Before we move to the experimental results, we highlight that the above discussion can be applied in a more general context to probe breaking of TR symmetry in crystals that belong to the $D_{3h}$ point group, and it could possibly be extended to other crystal symmetries. 

\subsection{Experimental results}

\begin{figure*}
\centering
\includegraphics[]{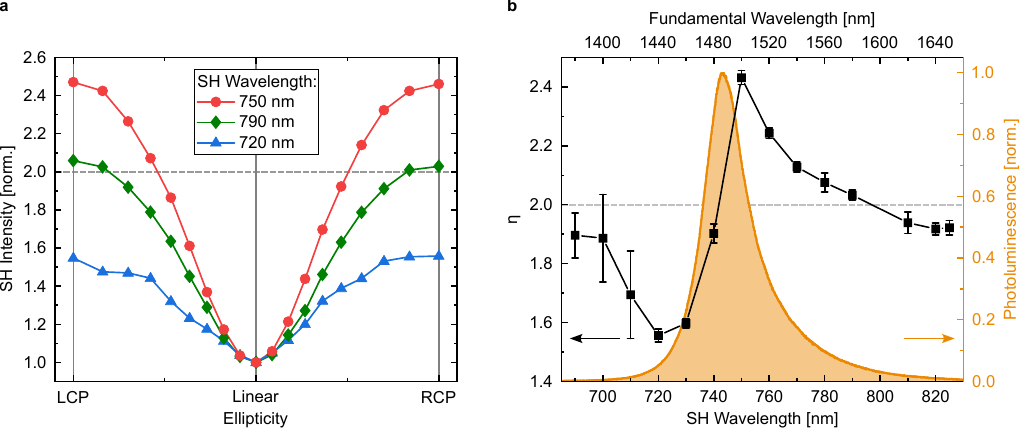}%
\caption{\label{fig:2}\textbf{Ellipticity dependence of SHG in monolayer {WSe}$_2$ across the $\mathbf{A\!\!:\!\!1s}$ exciton resonance} \textbf{a}, Normalized total emitted SH as a function of the ellipticity of the FB at different wavelengths across the ${A\!\!:\!\!1s}$ exciton. \textbf{b}, Wavelength-dependent ratio of SHG with circular to linear excitation (left axis, black squares). The grey dashed line indicates a ratio of 2. The orange curve is the normalized PL spectrum of the sample under investigation (emission at the ${A\!\!:\!\!1s}$ resonance).}
\end{figure*}

In order to demonstrate the features discussed in the previous section, we perform SHG experiments at different wavelengths (SH signal at \SI{690}{\nano\meter}-\SI{825}{\nano\meter}, corresponding to a FB in the range \SI{1380}{\nano\meter}-\SI{1650}{\nano\meter}) and scan across the ${A\!\!:\!\!1s}$ exciton resonance of a monolayer {WSe}$_2$ sample (see Methods for details on sample fabrication and characterization). In our experiments, we tune the ellipticity of the FB (from linear to circular) and detect the total SH intensity (see Methods for details on the experimental setup). Fig.\,\ref{fig:2}a shows the ellipticity dependent measurements for three selected wavelengths at resonance (\SI{750}{\nano\meter}), and away from resonance (\SI{790}{\nano\meter} and \SI{720}{\nano\meter}) with respect to the ${A\!\!:\!\!1s}$ exciton of our sample. Here, the FB power is kept constant at \SI{6}{\milli\watt} for all wavelengths. The curves are normalized (\textit{i.e.}, we set the SH intensity to 1 for linear excitation) and we paid particular attention to remove any possible contribution from two-photon PL (see Supplementary Information S4), which could alter the ratio of the circular/linear SH intensity. Clearly, this ratio is highly dispersive and can dramatically differ from two (\textit{i.e.}, the ratio expected from $D_{3h}$ symmetry), particularly in the case of resonant SHG. To further highlight this point, in Fig.\,\ref{fig:2}b we plot the wavelength dependence of $\eta$ (black squares) on top of the linear PL measured on the same sample (orange curve, see Methods and Supplementary Information S2 for detail on sample characterization). This ratio displays values both above and below 2 (horizontal dashed line) in correspondence of the exciton transition, with a shape that strongly resembles the derivative of the PL emission. While a detailed study of this shape is beyond the scope of this paper, we can simply observe that any nonlinear response can be decomposed, within the simple classical nonlinear harmonic oscillator model, as the product of linear susceptibilities (\textit{i.e.}, the Miller's rule and coefficient \cite{boyd2020nonlinear}). This can qualitatively explain why the dispersion reported in Fig.\,\ref{fig:2}b resembles the imaginary part of the linear dielectric constant in the same energy region, as measured for instance in differential reflectivity measurements that probe the exciton Rydberg states \cite{PhysRevLett.113.076802,RevModPhys.90.021001}. On the other hand, in the previous section we have demonstrated that while a ratio $> 2$ can be explained without considering the complex nature of the $\chi^{(2)}$ nonlinear tensor (equation\,(\ref{eq:3})), a ratio $< 2$ can only be explained by destructive interference between the VP and ED elements (equation\,(\ref{eq:4})). This leads to the conclusion that the VP and ED terms of the $\chi^{(2)}$ tensor are out of phase close to the ${A\!\!:\!\!1s}$ exciton resonance, in contrast to the hypothesis of previous reports \cite{Herrmann2023Nonlinear,Ho2020Measuring,Mouchliadis2021}. If we assume perfect constructive (destructive) interference between the ED and VP terms in this wavelength region, \textit{e.g.} the ED term is purely imaginary with a \SI{180}{\degree} phase shift at resonance (as predicted by the Lorentz model \cite{Lorentz1937}) while the VP term is purely real with no phase-shift, we can set the interference term in equation\,(\ref{eq:4}) to $\sin\Delta\varphi = \pm 1$ and thus calculate a value of the $\chi^{(2)}_{VP}$ of $\sim$\,\SI{41}{\pico\meter\per\volt} and $\sim$\,\SI{29}{\pico\meter\per\volt} at \SI{730}{\nano\meter} and \SI{750}{\nano\meter}, respectively (see Supplementary Information S5 for details), which correspond to $\sim$\,\SI{28}{\percent} and $\sim$\,\SI{18}{\percent} of the $\chi^{(2)}_{ED}$ at the same wavelengths. To the best of our knowledge, this work provides the first theoretical and experimental study of the complex values (real and imaginary part) of the $\chi^{(2)}_{VP}$ and their dispersion. 

\begin{figure*}
\centering
\includegraphics[]{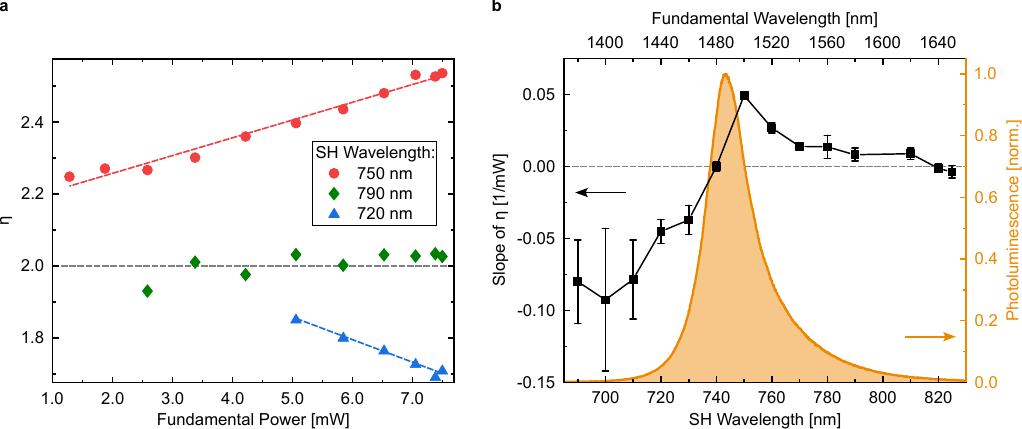}%
\caption{\label{fig:3}\textbf{Power dependence of the SHG ratio} \textbf{a}, The ratio of circular to linear SH depends linearly on the FB power for SH wavelengths of \SI{750}{\nano\meter} (red circles) and \SI{720}{\nano\meter} (blue triangles), while it is independent of power for \SI{790}{\nano\meter} (green diamonds). The grey dashed line marks a ratio of 2. \textbf{b}, Slope of the linear power dependencies (black squares, left axis) from \textbf{a} across the exciton resonance. The grey dashed line indicates a slope of 0, \textit{i.e.}, $\eta$ is power-independent. The orange curve is the normalized PL spectrum of the sample under investigation (emission at the ${A\!\!:\!\!1s}$ resonance).}
\end{figure*}

Finally, Fig.\,\ref{fig:3}a shows the power dependence of $\eta$ for three exemplary SH wavelengths, namely \SIlist{750;790;720}{\nano\meter}, while Fig.\,\ref{fig:3}b shows the slope (obtained by linear fitting of the curves in Fig.\,\ref{fig:3}a) of $\eta$ at wavelengths across the ${A\!\!:\!\!1s}$ exciton region of our sample. Here, the dotted horizontal line (slope $= 0$) corresponds to a ratio that is independent of power. In the absence of VP, the ratio should be indeed always equal to 2 and independent of power (equation\,(\ref{eq:3})), since the $\chi^{(2)}_{ED}$ probes an intrinsic property of the crystal. This is the case for wavelengths below the exciton resonance, see \textit{e.g.} Fig.\,\ref{fig:2}a and Fig.\,\ref{fig:3}b for wavelengths $>$\SI{780}{\nano\meter} (ratio $\sim\, 2$ and slope $\sim\, 0$). In contrast, in our experiments the VP term is a linear function of the FB power ($\chi^{(2)}_{VP} \sim\, I_\omega$), as it originates from the breaking and tuning of TR symmetry induced by optical Stark shift due to the off-resonant FB, as we already observed and discussed in Ref. \cite{Herrmann2023Nonlinear}. Note that the optical Stark effect neither involves nor requires a real excited state population, in contrast to the valley imbalance produced by two-photon absorption. In addition, this linear power dependence of $\eta$ confirms that the interference term in equation\,(\ref{eq:4}) dominates over the quadratic term $\frac{|\chi^{(2)}_{VP}|^2}{|\chi^{(2)}_{ED}|^2}$, because $|\chi^{(2)}_{VP}| \sim\, I_\omega$ and thus $|\chi^{(2)}_{VP}|^2 \sim\, I_\omega^2$. This is in agreement with the observation that $\frac{|\chi^{(2)}_{VP}|}{|\chi^{(2)}_{ED}|} \ll 1$, namely $\sim$\,\SI{28}{\percent} ($\sim$\,\SI{18}{\percent}) at \SI{730}{\nano\meter} (\SI{750}{\nano\meter}) and \SI{6}{\milli\watt} of average FB power, and thus $\frac{|\chi^{(2)}_{VP}|}{|\chi^{(2)}_{ED}|} \gg \frac{|\chi^{(2)}_{VP}|^2}{|\chi^{(2)}_{ED}|^2}$ in the power range of our experiments (Fig.\,\ref{fig:3}).

\subsection{Discussion}

In conclusion, we established a new method based on circular second harmonic generation to probe the valley degree of freedom in TMDs, and more generally to probe the breaking of time-reversal symmetry in crystals belonging to the $D_{3h}$ point group. We demonstrated that in such crystals the ratio between the circular and linear SH intensities can directly probe the valley induced nonlinear susceptibility $\chi^{(2)}_{VP}$ and we measured its dispersion in the wavelength region of the ${A\!\!:\!\!1s}$ exciton of a monolayer {WSe}$_2$ sample. From this, we could estimate values of $\chi^{(2)}_{VP} \sim\, $\SI{41}{\pico\meter\per\volt} and $\sim\, $\SI{29}{\pico\meter\per\volt} at \SI{730}{\nano\meter} and \SI{750}{\nano\meter} respectively, which correspond to a large fraction ($\sim\, $\SI{28}{\percent} and $\sim\, $\SI{18}{\percent}, respectively) of the intrinsic electric-dipole nonlinear response $\chi^{(2)}_{ED}$. Moreover, our approach provides direct access to the relative phase between the electric dipole and valley polarization generated second harmonics. This phase shift manifests itself in constructive and destructive interference across the ${A\!\!:\!\!1s}$ exciton resonance, in contrast to the commonly accepted assumption that the two terms are in phase at resonance. Finally, we have shown that while the $\chi^{(2)}_{ED}$ is independent of the excitation power, the $\chi^{(2)}_{VP}$ scales linearly with power in our experiments, confirming that here time-reversal symmetry is broken due to the coherent optical Stark effect. This work demonstrates the unique capabilities of nonlinear optics as a probe of broken time-reversal symmetry and of the valley degree of freedom in TMDs, and thus offers new insights for the development of nonlinear valleytronics, where parametric nonlinear optical processes can be used to probe valleys on ultrafast timescales and without perturbing the system.

\subsection{Online methods}

\subsubsection{Sample preparation and characterization}

We mechanically exfoliate a monolayer of {WSe}$_2$ from a bulk crystal (see Supplementary Information S1) onto PDMS and transfer it onto a transparent fused silica substrate. The monolayer nature of our sample is confirmed by optical contrast, PL, Raman and SHG (see Supplementary Information S2).

\subsubsection{Polarization resolved SHG}
For the SHG measurements we use a home-made multiphoton microscope, which we operate in transmission geometry (see Supplementary Information S2). The FB is generated by an optical parametric oscillator (Levante IR fs from APE), pumped by the output of an Yb doped mode locked laser (FLINT FL2-12, Light Conversion) with a repetition rate of \SI{76}{\mega\hertz} and pulse length of $\sim$\,\SI{100}{\femto\second}. This allows tuning of the FB from \SI{1300}{\nano\meter} to \SI{2000}{\nano\meter}. Before entering the microscope, a combination of halfwave-plate (AHWP05M-1600, Thorlabs) and quarterwave-plate (\#46-562, Edmund optics), both mounted in motorized rotation mounts (RSP05/M, Thorlabs), allows us to fully control the polarization state of the FB. Subsequently, the FB is focussed onto the sample by a x40 objective (LMM-40X-P01, Thorlabs) and the transmitted FB as well as the generated SH are collimated by a lens (C330TMD, Thorlabs). The transmitted FB is blocked by a shortpass filter (FESH0950, Thorlabs) and the SH is spectrally filtered by bandpass filters. Finally, we detect the SH with a Silicon avalanche-photo-diode (APD440A, Thorlabs) and lock-in amplifier (HF2LI, Zurich Instruments).

\section{Acknowledgments}
This work was funded by the German Research Foundation DFG (CRC 1375 NOA), project number 398816777 (subproject C4); the International Research Training Group (IRTG) 2675 “Meta-Active”, project number 437527638 (subproject A4); and by the Federal Ministry for Education and Research (BMBF) project number 16KIS1792 SINNER. Z.S. acknowledges the ERC-CZ program (project LL2101) from the Ministry of Education Youth and Sports (MEYS).


\end{document}



\section{Supplementary Information}
Valley Polarization-Electric Dipole Interference and Nonlinear Chiral Selection Rules in Monolayer {WSe}$_2$

\section{Author list}
Paul Herrmann$^1$, Sebastian Klimmer$^{1,2}$, Till Weickhardt$^1$, Anastasios Papavasileiou$^3$, Kseniia Mosina$^3$, Zden\v{e}k Sofer$^3$, Ioannis Paradisanos$^4$, Daniil Kartashov$^{5,6}$ and Giancarlo Soavi$^{1,6,\star}$

\section{Affiliations}
\noindent
$^1$Institute of Solid State Physics, Friedrich Schiller University Jena, Helmholtzweg 5, 07743 Jena, Germany
\newline
$^2$ARC Centre of Excellence for Transformative Meta-Optical Systems, Department of Electronic Materials Engineering, Research School of Physics, The Australian National University, Canberra, ACT, 2601, Australia
\newline
$^3$Department of Inorganic Chemistry, University of Chemistry and Technology, Technicka 5, Prague, 166 28 Czech Republic
\newline
$^4$Institute of Electronic Structure and Laser, Foundation for Research and Technology, N. Plastira 100, Vassilika Vouton, 70013 Heraklion, Crete, Greece
\newline
$^5$Institute of Optics and Quantum Electronics, Friedrich Schiller University Jena, Max-Wien-Platz 1, 07743 Jena, Germany
\newline
$^6$Abbe Center of Photonics, Friedrich Schiller University Jena, Albert-Einstein-Straße 6, 07745 Jena, Germany
\newline
$^{\star}$ giancarlo.soavi@uni-jena.de

\maketitle

\maketitle

\section{S1 Synthesis of bulk Tungsten Diselenide crystals}

The synthesis of {WSe}$_2$ was performed by chemical vapor transport in a quartz glass ampoule from Tungsten (\SI{99.999}{\percent}, -100 mesh, China Rhenium Co., Ltd, China) and Selenium (\SI{99.9999}{\percent} granules 1-6mm, Wuhan Xinrong New Material Co., Ltd., China) in the stochiometric amount corresponding to 100g of {WSe}$_2$. In addition, excess of 2 at.\% of selenium, 0.5g of {SeCl}$_4$ (99.9\%, rough crystalline powder, Strem, USA) and 0.5g of iodine (99.9\%, granules, Fisher Scientific, USA) were added in a glovebox to the ampoule (\SI{50}{}x\SI{250}{\milli\meter}, wall thickness \SI{3}{\milli\meter}) and the ampoule was sealed by an oxygen-hydrogen welding torch under high vacuum (under \SI{1e-3}{\pascal}) using a diffusion pump with liquid nitrogen trap. The sealed ampoule was first placed in a muffle furnace and heated to \SI{500}{\celsius} for 25 hours, \SI{600}{\celsius} for 50 hours and finally \SI{800}{\celsius} for 50 hours. The heating and cooling rate was \SI{1}{\celsius\per\minute}. The ampoule with formed powder {WSe}$_2$ was placed in a two zone horizontal furnace. First, the growth zone was heated to \SI{1000}{\celsius} and the source zone to \SI{800}{\celsius}. After 2 days the thermal gradient was reversed, as the source zone was kept at \SI{1000}{\celsius} while the growth zone was kept at \SI{900}{\celsius} for 10 days. During the cooling the thermal gradient was reversed for 2 hours, in order to remove the transport medium and volatile compounds. The ampoule was opened in an argon filled glovebox. 

\section{S2 Sample fabrication, characterization and setup}

\begin{figure*}[h!]
\centering
\includegraphics[]{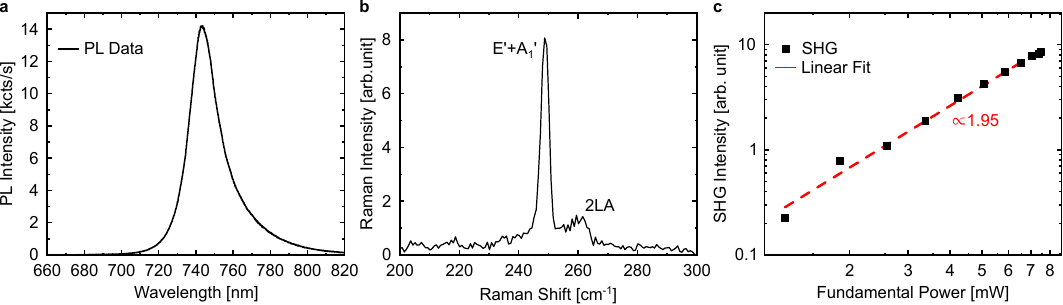}%
\caption{\label{fig:S1}\textbf{Characterization of monolayer {WSe}$_2$} \textbf{a}, Photoluminescence spectrum of the {WSe}$_2$ sample. \textbf{b}, Raman spectrum showing the characteristic {E'}, {A$_{1}$'} and {2LA} modes. \textbf{c}, Double-logarithmic plot of the SHG intensity (black squares) over the fundamental power. The linear fit (red dashed line) shows a slope of \SI{1.95}{}.}
\end{figure*}

The sample was prepared by mechanical exfoliation of a bulk crystal produced with the procedure described in Section S1. To confirm the monolayer nature of our sample, we measured the PL (Fig.\,\ref{fig:S1}a) under excitation with a CW laser ($\lambda_{exc}=\,$\SI{532}{\nano\meter}). The strong PL is characteristic for monolayers, as the bandgap shifts from direct in the monolayer case to indirect in the bi-layer to bulk case \cite{TMD_bandgap}. We observe the PL maximum at $\sim$\,\SI{743}{nm}, corresponding to the A exciton in {WSe}$_2$ in agreement with literature \cite{Zeng2013}. In addition, we measure the Raman spectrum of our sample (Fig.\,\ref{fig:S1}b), which shows the for monolayer WSe$_2$ characteristic degenerate E' and A$_1$' modes at $\sim$\,\SI{249}{\per\centi\meter} and a second order 2LA mode at $\sim$\,\SI{260}{\per\centi\meter} \cite{del2016atypical}. Finally, we performed power dependent SHG measurements (Fig.\,\ref{fig:S1}c). The data shows a power scaling with an exponent of nearly 2, which is characteristic for second order nonlinear processes. Due to crystal symmetry, SHG in TMDs is possible only in odd number of layers  \cite{Säynätjoki2017}. Since we can safely rule out a tri-layer from the PL spectroscopy and optical contrast, we can further confirm the monolayer nature of our {WSe}$_2$ sample.

\begin{figure*}[h!]
\centering
\includegraphics[]{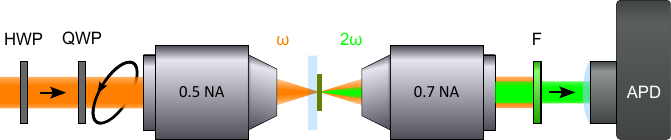}%
\caption{\label{fig:S3}\textbf{Multiphoton microscopy setup} Experimental setup with the half-wave (HWP) and quarter-wave plate (QWP) to control the polarization state of the FB ($\omega$, orange). The FB is focused onto the sample and the transmitted FB as well as the generated SH ($2\omega$, green) are collimated. Subsequently, the SH is filtered (F) and detected by a Silicon avalanche photo-diode (APD).}
\end{figure*}

For the polarization-resolved SHG measurements we used a home-made multiphoton microscope (Fig.\,\ref{fig:S3}). A combination of a half- and quarter-wave plate gives us complete control over the polarization state of the incident fundamental beam (FB). The generated SH signal is collected in transmission geometry, spectrally filtered and subsequently detected with a silicon avalanche photo-diode.

\section{S3 Second harmonic generation in TMD monolayers}

In this section we first revise the intrinsic second order nonlinear response, \textit{i.e.}, in the absence of a valley polarization (VP), of a TMD monolayer. Following that we generalize the response to include the valley susceptibility, carefully analyzing the resulting modified second order nonlinear polarization. Finally, we derive the ratio $\eta$ of circular to linear SHG depending on the valley susceptibility.

\subsection{S3.1 Intrinsic response}

As mentioned in the main text, TMD monolayers in absence of VP belong to the $D_{3h}$ point group. This point group has four non-independent elements of the second order nonlinear susceptibility, namely $\chi^{(2)}_{ED}=\chi^{(2)}_{xxx}=-\chi^{(2)}_{xyy}=-\chi^{(2)}_{yyx}=-\chi^{(2)}_{yxy}$, where $x(y)$ refers to armchair (zig-zag) axis of the crystal. We label this the ED (electric dipole or intrinsic) response of a TMD monolayer, as it describes its crystal (broken space-inversion) symmetry. In the contracted notation the second order nonlinear polarization is then given by:

\begin{equation} \label{eq:1}
    \boldsymbol{P^{(2)}} = 
    \begin{pmatrix}
        P_{x}^{(2)}\\
        P _{y}^{(2)}
    \end{pmatrix}
    = \epsilon_0
    \begin{pmatrix}
        \chi_{ED}^{(2)} & -\chi_{ED}^{(2)} & 0\\
        0 & 0 & -\chi_{ED}^{(2)}
    \end{pmatrix}
    \cdot
    \begin{pmatrix}
        E_{x}^{2}\\
        E_{y}^2\\
        2 E_{x} E_{y}
    \end{pmatrix}
    = \epsilon_0
    \begin{pmatrix}
        \chi_{ED}^{(2)} \cdot (E_x^2 - E_y^2)\\
        -2\chi_{ED}^{(2)} \cdot E_x E_y
    \end{pmatrix}
\end{equation}

where we neglect the $z$-component since the light propagates along the $z$-axis and the TMD monolayer has vanishing thickness.

\subsection{S3.2 Adding the VP-induced response}

In the presence of a VP, the symmetry of the TMD monolayer is reduced from $D_{3h}$ to $C_{3h}$ \cite{Herrmann2023Nonlinear}, adding the following four non-independent elements to the total susceptibility: $\chi^{(2)}_{VP}=\chi^{(2)}_{yyy}=-\chi^{(2)}_{yxx}=-\chi^{(2)}_{xxy}=-\chi^{(2)}_{xyx}$, which we label as the VP susceptibility. This VP susceptibility is a sum of the contributions from the $\pm K$ valleys, $\chi^{(2)}_{VP}=\chi^{(2)}_{VP}(K)+\chi^{(2)}_{VP}(-K)$, however the $\pm K$ valley can only be addressed by left/right circularly polarized light. Furthermore, the $\pm K$ valleys are connected by time-reversal symmetry, resulting in a valley susceptibility of equal magnitude but opposite sign: $\chi^{(2)}_{VP}(K)= - \chi^{(2)}_{VP}(-K)$. Therefore the total second order nonlinear polarization in presence of a VP can be written as:

\begin{equation} \label{eq:2}
    \boldsymbol{P^{(2)}} = 
    \begin{pmatrix}
    P_x^{(2)} \\
    P_y^{(2)}
    \end{pmatrix}
    = \epsilon_0
    \begin{pmatrix}
    \chi_{ED}^{(2)}(E_x^2-E_y^2) - 2 \tau \chi_{VP}^{(2)} E_x E_y\\
    - 2 \chi_{ED}^{(2)} E_x E_y -  \tau \chi_{VP}^{(2)} (E_x^2 - E_y^2)
    \end{pmatrix}.
\end{equation}

Note the pre-factor $\tau$ in front of the VP term, due to the aforementioned selection rule and symmetry. We can gain better insights into this by switching from the linear $\boldsymbol{e_x}/\boldsymbol{e_y}$ basis to the circular $\boldsymbol{\sigma_+}/\boldsymbol{\sigma_-}$ basis, using the definition $\boldsymbol{\sigma_\pm} = \frac{1}{\sqrt{2}}(\boldsymbol{e_x} \pm i \boldsymbol{e_y})$. In this circular basis the electric field can be written as $\boldsymbol{E}=E_+ \boldsymbol{e_+} + E_- \boldsymbol{e_-}$ with amplitudes 

\begin{equation} \label{eq:3}
    E_{\pm}=\frac{1}{\sqrt{2}}(E_x \pm i E_y) 
\end{equation}
    
of the left/right circular components. Analogously the second order nonlinear polarization can be represented in this circular basis ($\boldsymbol{P^{(2)}}=P_+^{(2)} \boldsymbol{e_+} + P_-^{(2)} \boldsymbol{e_-}$) with the left/right circular amplitudes $P_{\pm}^{(2)} = \frac{1}{\sqrt{2}}(P_x^{(2)} \pm i P_y^{(2)})$. Therefore, conversion of the polarization into the circular basis leads to:

\begin{equation} \label{eq:4}
    \boldsymbol{P^{(2)}} = 
    \begin{pmatrix}
    P_+ \\
    P_-
    \end{pmatrix}
    = \epsilon_0 \frac{1}{\sqrt{2}}
    \begin{pmatrix}
    \chi_{ED}^{(2)}(E_x^2 - E_y^2 - 2 i E_x E_y) - i \tau \chi_{VP}^{(2)}(E_x^2 - E_y^2 - 2 i E_x E_y)\\
    \chi_{ED}^{(2)}(E_x^2 - E_y^2 + 2 i E_x E_y) + i \tau \chi_{VP}^{(2)}(E_x^2 - E_y^2 + 2 i E_x E_y)
    \end{pmatrix}.
\end{equation}

We can now substitute $E_{\pm}^2= \frac{1}{2}(E_x^2 - E_y^2 \pm 2i E_x E_y)$ from equation (\ref{eq:3}) and finally obtain:

\begin{equation} \label{eq:5}
    \boldsymbol{P^{(2)}} =
    \begin{pmatrix}
    P_+ \\
    P_-
    \end{pmatrix}
    = \epsilon_0 \sqrt{2}
    \begin{pmatrix}
    (\chi_{ED}^{(2)} - i \tau \chi_{VP}^{(2)}) E_-^2\\
    (\chi_{ED}^{(2)} + i \tau \chi_{VP}^{(2)}) E_+^2
    \end{pmatrix}
    = \epsilon_0 \sqrt{2}
    \begin{pmatrix}
    (\chi_{ED}^{(2)} + i \chi_{VP}^{(2)}) E_-^2\\
    (\chi_{ED}^{(2)} + i \chi_{VP}^{(2)}) E_+^2
    \end{pmatrix}.
\end{equation}

Note that in the last step we resolved $\tau$ according to $\tau(E_{\pm})=\pm 1$, as discussed above. 

\subsection{S3.3 Measuring VP}

Now we focus on the total emitted SHG for linear and circular polarization. For linear excitation the electric field is given by $\boldsymbol{E}=E_0 \cdot(\cos{(\theta)} \, \boldsymbol{e_x}+\sin{(\theta)} \, \boldsymbol{e_y})$, where $\theta$ defines the angle between the AC-axis of the crystal and the polarization-axis of the FB. The resulting SH is then proportional to the absolute square of the second order nonlinear polarization ($I_{SHG} \propto |\boldsymbol{P^{(2)}}|^2$). In the linear case, no VP is induced and thus the second order nonlinear polarization is given by equation (\ref{eq:1}). In this case, the total emitted SH intensity 

\begin{equation} \label{eq:6}
    I_{lin}(2\omega) \propto |\chi_{ED}^{(2)}|^2 |E_0|^4
\end{equation}

is proportional to the fourth power of the fundamental field, \textit{i.e.}, to the square of the intensity of the fundamental field ($|E_0|^4=I(\omega)^2$) and of the ED susceptibility $|\chi_{ED}^{(2)}|^2$. However, as discussed in the main text, changing the excitation to circular induces a VP. We take this into account by using equation (\ref{eq:5}). For circularly polarized light the total emitted SH 

\begin{equation} \label{eq:7}
    I_{circ}(2\omega) \propto 2 \cdot |\chi_{ED}^{(2)}+i \chi_{VP}^{(2)}|^2 |E_0|^4
\end{equation}

is still proportional to the square of the fundamental intensity, but with a pre-factor of 2 and the additional term $|\chi_{ED}^{(2)}|^2 \rightarrow |\chi_{ED}^{(2)}+i \chi_{VP}^{(2)}|^2$. While the pre-factor 2 is not related to the VP, and was observed previously in Ref. \cite{PhysRevB.98.115426}, the VP elements of the nonlinear susceptibility allow to directly probe the broken TR symmetry. Taking the ratio of SH intensities with circular to linear excitation we obtain:

\begin{equation} \label{eq:8}
    \eta := \frac{I_{circ}(2\omega)}{I_{lin}(2\omega)} = 2 \cdot \frac{|\chi_{ED}^{(2)}+i \chi_{VP}^{(2)}|^2}{|\chi_{ED}^{(2)}|^2}
\end{equation}

where we can observe that the ratio is indeed 2 in the absence of a VP ($\chi_{VP}^{(2)}=0$) and otherwise will differ from 2. Carefully resolving the fraction of the absolute squares and considering the complex nature of the elements of the nonlinear susceptibilities, we end up with:

\begin{equation} \label{eq:9}
    \eta = 2 \cdot \left[1 + \frac{|\chi^{(2)}_{VP}|^2}{|\chi^{(2)}_{ED}|^2} + \frac{1}{|\chi^{(2)}_{ED}|^2} \Big\{\Re\big(\chi^{(2)}_{VP}\big) \cdot \Im\big(\chi^{(2)}_{ED}\big) - \Im\big(\chi^{(2)}_{VP}\big) \cdot \Re\big(\chi^{(2)}_{ED}\big)\Big\} \right]
\end{equation}

where the symbols $\Re$ and $\Im$ represent real and imaginary parts, respectively. 
Finally, we can rewrite the last term by representing the susceptibilites as $\chi_{ED/VP}^{(2)}=|\chi_{ED/VP}^{(2)}| \cdot e^{i \varphi_{ED/VP}}$ as a product of their amplitude $|\chi_{ED/VP}^{(2)}|$ and phase $\varphi_{ED/VP}$: 

\begin{equation} \label{eq:10}
    \frac{1}{|\chi^{(2)}_{ED}|^2} \Big\{\Re\big(\chi^{(2)}_{VP}\big) \cdot \Im\big(\chi^{(2)}_{ED}\big) - \Im\big(\chi^{(2)}_{VP}\big) \cdot \Re\big(\chi^{(2)}_{ED}\big)\Big\} = \frac{|\chi^{(2)}_{VP}|}{|\chi^{(2)}_{ED}|} \cdot \sin{\Delta\varphi}
\end{equation}

where $\Delta\varphi = \varphi_{ED}-\varphi_{VP}$. 

\section{S4 Removal of TP-PL from SHG}

\begin{figure*}
\centering
\includegraphics[]{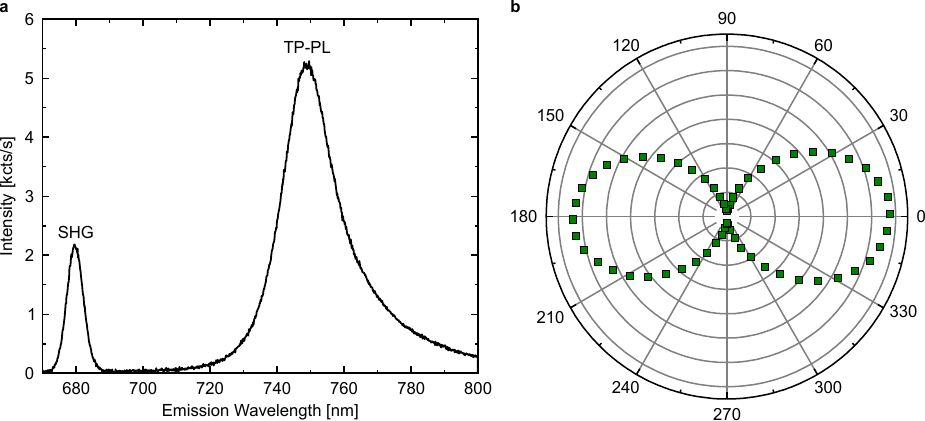}%
\caption{\label{fig:S2}\textbf{Two-photon PL and SHG in monolayer {WSe}$_2$} \textbf{a}, SHG and TP-PL for excitation wavelength of \SI{1360}{\nano\meter}. \textbf{b}, Polarization scan of the emitted SH by rotating a linear polarizer in front of the detector.}
\end{figure*}

When scanning the SH emission wavelength across the exciton $A:1s$ resonance, the signal may overlap with two-photon-photoluminescence (TP-PL, see Fig.\,\ref{fig:S2}a). Thus, in our resonant SHG measurements the TP-PL must be properly subtracted. Since the valley coherence for two-photon-excitation at room temperature in monolayer {WSe}$_2$ is expected to be 0 (see \textit{e.g.} Ref. \cite{Herrmann2023Nonlinear}), TP-PL is unpolarized and does not depend on the polarization of the excitation laser. Therefore we can assume that the emitted TP-PL is equal in the cases of linear and circular excitation. The emitted SH for linear excitation is linear, see Fig.\,\ref{fig:S2}b, providing a pathway to suppress the SH, by means of a linear polarizer oriented perpendicular to the SH, and measuring the residual TP-PL. Hence, after recording the total SH for circular and linear excitation, we set the excitation to linear and insert a linear polarizer in the correct orientation to suppress the SH and record the TP-PL. We follow this procedure for all excitation wavelengths and powers. Finally, in the data analysis we subtract the TP-PL from the SH.

\section{S5 Calculation of second order nonlinear susceptibility}

To calculate the intrinsic second order nonlinear optical susceptbility we use the following formula from Ref. \cite{Rosa2018} for SHG in transmission geometry:

\begin{equation} \label{eq:11}
    |\chi_S^{(2)}| = \sqrt{\frac{P_{SH}}{P_{FF}^2} \cdot \frac{c \epsilon_0 f r^2 t_{FWHM}\lambda^2}{64 \sqrt{2}\pi S} \cdot \frac{(1+n)^6}{n^3}}
\end{equation}

with the powers $P_{FF}$ and $P_{SH}$ of the fundamental and emitted SH, respectively, the speed of light $c$ and the vacuum electric permittivity $\epsilon_0$. $f=\;$\SI{76}{\mega\hertz} is the repetition rate of the pump laser, $r\approx\,$\SI{1.85}{\micro\meter} is the radius of the focal spot, $t_{FWHM}\approx\,$\SI{200}{\femto\second} is the FWHM of the pulse and $\lambda$ (\SI{1380}{\nano\meter} - \SI{1650}{\nano\meter}) is the wavelength of the FB. $S=\,$0.94 is the shape factor of a Gaussian pulse and $n\approx\,$1.44 is the refractive index of the fused silica substrate in the wavelength range of the FB. In order to calculate the VP susceptibility, we analyze equation (\ref{eq:8}) and (\ref{eq:9}). As mentioned in the main text, we assume perfect constructive and destructive interference between the ED and VP contributions for SH wavelengths of \SI{750}{\nano\meter} and \SI{730}{\nano\meter}, respectively. Therefore the ratio is then given by:

\begin{equation} \label{eq:12}
    \eta = 2 \cdot \left[1 + \frac{|\chi^{(2)}_{VP}|^2}{|\chi^{(2)}_{ED}|^2} \pm \frac{|\chi^{(2)}_{VP}|}{|\chi^{(2)}_{ED}|} \right]
\end{equation}

where the $\pm$ corresponds to constructive/destructive interference. Solving this equation we retrieve the ratio of the absolute values of the susceptibilities as:

\begin{equation} \label{eq:13}
    \frac{|\chi^{(2)}_{VP}|}{|\chi^{(2)}_{ED}|} = \mp \frac{1}{2} \pm \sqrt{\frac{2 \cdot \eta -3}{4}}
\end{equation}

with the $\mp$ coming from the constructive/destructive interference, while the $\pm$ appears since equation (\ref{eq:12}) is a quadratic formula. Two points can be noted: (1) $\frac{|\chi^{(2)}_{VP}|}{|\chi^{(2)}_{ED}|}$ has to be positive; (2) in general there are two solutions. For a SH wavelength of \SI{750}{\nano\meter} (constructive interference), with $\eta=\;$\SI{2.43}{} we obtain $\frac{|\chi^{(2)}_{VP}|}{|\chi^{(2)}_{ED}|}=\;$\SI{0.18}{} and $\frac{|\chi^{(2)}_{VP}|}{|\chi^{(2)}_{ED}|}=\;$\SI{-1.18}{}. However, the second solution is un-physical, as it is negative. For a SH wavelength of \SI{730}{\nano\meter} (destructive interference), with $\eta=\;$\SI{1.59}{} we obtain two possible solutions: $\frac{|\chi^{(2)}_{VP}|}{|\chi^{(2)}_{ED}|}=\;$\SI{0.71}{} and $\frac{\chi^{(2)}_{VP}|}{|\chi^{(2)}_{ED}|}=\;$\SI{0.29}{}. In the main text we present only the latter as the lower limit.  

\section{S6 Definition of error bars}

The raw data for each measurement is taken as follows: for each data point we average three times, and define the mean as the measured value and the standard deviation as the error. This way we consider the optical noise, \textit{i.e.}, fluctuations in either the fundamental or SH. Note that we do not plot the error bars corresponding to this standard deviation in the main text in Fig. 2a, because the error bars are smaller than the symbols. However, as we detect the signal electrically, we need to also account for the electrical noise of the lock-in setup, including APD, optical chopper and lock-in detector. We retrieve the electrical noise \textit{via} a dark measurement, where we block the detector and record the signal. The average of this dark signal is then the electrical noise. Every SH data point, where the signal is smaller than the electrical noise is disregared, as it can not be differentiated from the electrical noise.

For the wavelength dependence of the ratio circular to linear SHG (see Fig.\,2b in the main text), we fit the ellipticity $\epsilon$ dependent SHG data (including the error bars given by the standard deviation) with the following function:

\begin{equation}
    I(\epsilon) = A \sin{(\pi\epsilon)}^2 + I_{lin}
\end{equation}
\textit{i.e.}, the square of a sinusoidal function with the amplitude $A$ plus the offset $I_{lin}$, which corresponds to the intensity for linear excitation. The intensity for circular excitation is then given by the sum of the amplitude and the offset ($I_{circ} = A + I_{lin}$). By fitting the data we retrieve the parameters ($A$,$I_{lin}$) as well as their standard errors ($\Delta A$,$\Delta I_{lin}$).  Although this directly gives us the error of SHG intensity for linear excitation ($\Delta I_{lin}$), for the ratio we need to take the formula for error propagation:

\begin{equation} \label{eqn:error}
    \Delta \eta \approx \frac{\partial \eta}{\partial A} \cdot \Delta A + \frac{\partial \eta}{\partial I_{lin}} \cdot \Delta I_{lin} = \frac{1}{I_{lin}} \cdot \Delta A + \frac{A}{I_{lin}^2} \cdot \Delta I_{lin}
\end{equation}
to calculate the error bars in Fig.\,2b of the main text. 

We measure the SH power dependencies four times, then take the average

\begin{equation}
    SH = \frac{1}{4} \sum_{i=1}^4 SH_i
\end{equation}

and propagate the error according to:

\begin{equation}
    \Delta SH \approx \frac{1}{4} \sum_{i=1}^4 SH_i \cdot \Delta SH_i 
\end{equation}

derived from the formula for error propagation (see equation (\ref{eqn:error})). Especially for the power dependent measurements, taking the electrical noise is of great significance, as for certain wavelengths the signal can be very low, so that a big portion of data lies within the electrical noise and needs to be disregarded.
After averaging, propagating the error and disregarding the data smaller than the electrical noise, we plot the power dependencies in Fig.\,3a of the main text. Note that the error bars are smaller than the symbols and therefore not plotted. We fit the power dependencies (including error) with a linear function, and obtain the slopes together with their standard errors, which we set as the error bars, see Fig.\,3b of the main text.
